\documentclass[aps,prb,reprint,amsmath,amssymb,floatfix,longbibliography]{revtex4-2}
\usepackage{physics,xcolor,bm,microtype,graphicx,hyperref}
\hypersetup{colorlinks=true,allcolors=blue}

\begin{document}
\title{Quantum geometry and magnon Hall transport in an altermagnet}
\author{Erlend Sylju{\aa}sen}
\affiliation{Center for Quantum Spintronics, Department of Physics, Norwegian University of Science and Technology, NO-7491 Trondheim, Norway}
\author{Alireza Qaiumzadeh}
\affiliation{Center for Quantum Spintronics, Department of Physics, Norwegian University of Science and Technology, NO-7491 Trondheim, Norway}
\author{Asle Sudb{\o}}
\affiliation{Center for Quantum Spintronics, Department of Physics, Norwegian University of Science and Technology, NO-7491 Trondheim, Norway}
\date{\today}
\begin{abstract}
We compute magnon Hall conductivities in a minimal model of a two-dimensional altermagnet. To do so, we derive an analytic expression for the relevant quantum geometric tensor describing two-band bosonic Bogoliubov Hamiltonians, providing insight into the geometric, topological, and transport properties. The magnon thermal Hall and spin Nernst conductivities are shown to directly depend on the altermagnetic parameter, which may serve as an experimental probe of altermagnetism.
\end{abstract}
\maketitle
\section{Introduction}
Altermagnets have attracted significant attention because of their potential applications in spintronics and other fields \cite{smejkal_2022}. Specifically, the combination of compensated collinear magnetic order and a large nonrelativistic momentum-dependent spin-split band structure makes them particularly interesting and distinct from conventional antiferromagnets and ferromagnets \cite{Noda_2016, Hayami_2019}. Despite promising applications and numerous material candidates, definitive experimental observations remain limited \cite{fedchenko_2024, amin_2024, jiang_2024, zhang_2024}. This underscores the need for effective probes of altermagnetism. 

Thermomagnetoelectric phenomena are crucial for designing energy-efficient devices, thermal management systems, and novel sensors. Various thermomagnetoelectric longitudinal and transverse signals can be used to indirectly distinguish topological and magnetic states of quantum systems \cite{Li_2020,Wang_2023,qauimzadeh_brehm_2025}. In magnetic insulators, a temperature gradient can drive magnon-mediated spin and/or heat currents with transverse components from Berry curvature induced anomalous velocity \cite{matsumoto_2011}. We emphasize that the magnon Hall effects discussed here originate from the antisymmetric off-diagonal components of the response functions, and are thus distinct from other transverse responses in altermagnets that arise from the symmetric off-diagonal components \cite{Hernandez_2021,hongxin_2023,Zarzuela_2025,marmodoro_2024,sun_2025}.

The main theoretical distinction between magnon Hall effects and their fermionic counterparts is the bosonic character of magnon excitations. Bose-Einstein statistics prevent quantized magnon Hall conductivities in the zero temperature limit. Furthermore, magnon Hamiltonians can have particle non-conserving terms that necessitate a bosonic Bogoliubov-Valatin transformation. This is not a unitary transformation \cite{colpa_1978}, which leads to a modified quantum geometric tensor \cite{eckardt_2024}.

This paper is organized as follows. In Sec. \ref{sec:two_band_bosonic_bogoliubov_hamiltonian}, we develop a framework for computing magnon Hall conductivities in generic two-band bosonic Bogoliubov Hamiltonians. In particular, we identify the pseudo-unitary group structure and use this to obtain a convenient expression for the relevant quantum geometric tensor. In Sec. \ref{sec:lieb_lattice_altermagnet}, we apply this framework to a Lieb lattice altermagnet and compute magnon Hall conductivities. Finally, we summarize our results in Sec. \ref{summary}.

\section{Two-Band Bosonic Bogoliubov Hamiltonian} \label{sec:two_band_bosonic_bogoliubov_hamiltonian}
We consider a two-band bosonic Bogoliubov Hamiltonian that may originate from mapping a spin Hamiltonian describing an antiferromagnet with two magnetic sublattices to bosonic operators. The Hamiltonian reads 
\begin{align} \label{eq:general_hamiltonian}
    \mathcal{H} &= \sum_{\bm{k}} \bm{\varphi}^\dagger_{\bm{k}} H_{\bm{k}} \bm{\varphi}_{\bm{k}}
    \\
    &= \sum_{\bm{k}}
    \begin{pmatrix}
        b_{1, \bm{k}}^\dagger & b_{2, -\bm{k}}
    \end{pmatrix}
    \begin{pmatrix}
        f_{\bm{k}} & r_{\bm{k}}
        \\
        r^*_{\bm{k}} & g_{\bm{k}}
    \end{pmatrix}
    \begin{pmatrix}
        b_{1, \bm{k}} \\ b_{2, -\bm{k}}^\dagger
    \end{pmatrix}, \notag
\end{align}
where the summation over crystal momentum $\bm{k}$ is restricted to the first Brillouin zone. For brevity, we suppress explicit dependence on $\bm{k}$ and write partial derivatives as $\partial_i = \partial_{k_i}$. The single-particle Hamiltonian is $H$ with $r = r_1 + i r_2$ and $f, g, r_1, r_2 \in \mathbb{R}$ ensuring its Hermiticity.  We focus on the nontrivial case where $\abs{r_{1}} + \abs{r_{2}} \neq 0$. The Nambu array $\bm{\varphi}$ contains a bosonic annihilation operator $b_1$ and a bosonic creation operator $b_2^\dagger$ that obeys commutation relations of the form
\begin{align} \label{eq:commutation_relations}
    &[\varphi_n, \varphi_m] = 0, &[\varphi_n, \varphi_m^\dagger] = (\sigma_z)_{nm} \, 1_F,
\end{align}
where $n, m \in \{1, 2\}$, $\sigma_z$ is the Pauli-$z$ matrix, and $1_F$ denotes the Fock space identity. In the following, we introduce the bosonic Bogoliubov-Valatin transformation as outlined in Ref. \cite{colpa_1978} to diagonalize the Hamiltonian. We then emphasize the pseudo-unitary group structure and derive an expression for the relevant quantum geometric tensor.

\subsection{Bosonic Bogoliubov-Valatin transformation}
We introduce new operators $\bm{\psi} = (d_1, d_2^\dagger)^\mathsf{T}$ defined as linear combinations of the original ones via $\bm{\varphi} = T \bm{\psi}$, where $T$ is a $2 \times 2$ complex matrix. Applying this transformation in Eq. \eqref{eq:general_hamiltonian}, we require that
\begin{equation} \label{eq:quasiparticle_energies}
    T^\dagger H T = E
\end{equation}
becomes a diagonal matrix holding the excitation energies, $E = \mathrm{diag}(E_+, E_-)$. Note that $E_\pm = E_\pm(\pm \bm{k})$ due to the signs of $\bm{k}$ in Eq. \eqref{eq:general_hamiltonian}. We also require that the new operators remain bosonic, satisfying the commutation relations in Eq. \eqref{eq:commutation_relations}, which leads to the constraint
\begin{equation} \label{eq:para_unitary_constraint}
    T^\dagger \sigma_z T = \sigma_z. 
\end{equation}
This confirms that $T$ is not unitary and a matrix satisfying this condition is commonly referred to as being pseudo-unitary or paraunitary \cite{colpa_1978}. We assume both thermodynamic and dynamic stability, which implies the existence of a pseudo-unitary matrix $T$ that gives positive and purely real excitation energies. Reference \cite{colpa_1978} shows that $H$ being positive definite is a sufficient condition for these assumptions. Therefore, we assume $\tr H > 0$ and $\det H > 0$. 

Substituting Eq. \eqref{eq:para_unitary_constraint} into Eq. \eqref{eq:quasiparticle_energies} gives
\begin{equation} \label{eq:diagonalization_dynamical}
    T^{-1} D T = \Lambda,
\end{equation}
where $D = \sigma_z H$ is named the dynamical matrix and $\Lambda = \sigma_z E$ contains the excitation energies. Therefore, diagonalizing the dynamical matrix yields the transformation matrix and the excitation energies. Another hint of why $D$ is the important matrix to consider is obtained by computing the time evolution of the Nambu array,
\begin{equation}
    \dv{t} \bm{\varphi}(t) = \frac{i}{\hbar}[\mathcal{H}, \bm{\varphi}(t)] = -\frac{i}{\hbar}D \bm{\varphi}(t).
\end{equation}
Here, the Heisenberg picture is defined by $\bm{\varphi}(t) = e^{i\mathcal{H}t/\hbar}\bm{\varphi} e^{-i\mathcal{H}t/\hbar}$ and the presence of $D$ on the right-hand side indicates that $D$ is the generator of dynamics \cite{sato_2019, flynn_2023}. Moreover, it is the eigenvalues and eigenvectors of $D$ that explicitly enter the expressions for the transport coefficients to be computed later. 

The dynamical matrix is non-Hermitian ($D^\dagger \neq D$), but can be shown to be Hermitian with respect to another indefinite inner product \cite{sato_2019}, which will be labeled pseudo-Hermiticity in the following. From Eq. \eqref{eq:para_unitary_constraint}, it follows that the columns of $T$ are normalized according to $\langle v_n, \sigma_z v_m \rangle = (\sigma_z)_{nm}$, where $v_n, v_m$ are columns of $T$ with $n, m \in \{1, 2\}$. As in Ref. \cite{sato_2019}, we define the Krein inner product of two vectors $u, w \in \mathbb{C}^2$ as $\langle u, w \rangle_{\sigma_z} = \langle v, \sigma_z w \rangle$ with $\sigma_z$ being the Krein weight of the Krein inner product. Moreover, we define the pseudo-adjoint of a linear operator $A$ by the relation $\langle u, A w \rangle_{\sigma_z} = \langle A^{\#} u, w \rangle_{\sigma_z}$. Using the Hermiticity of $\sigma_z$ gives 
\begin{equation}
	A^{\#} = \sigma_z A^\dagger \sigma_z.
\end{equation}
Following Ref. \cite{sato_2019}, we further introduce pseudo-unitarity as $A^{\#} = A^{-1}$, pseudo-Hermiticity as $A^{\#} = A$, and a pseudo-orthogonal projection as $A^{\#} = A = A^2$. The transformation matrix $T$ is observed to be pseudo-unitary using Eq. \eqref{eq:para_unitary_constraint} and the dynamical matrix $D$ is seen to be pseudo-Hermitian using the Hermiticity of $H$.

We note that Eq. \eqref{eq:para_unitary_constraint} defines the (indefinite) pseudo-unitary group with signature $(1, 1)$, meaning that $T \in \text{U}(1,1)$. The Krein inner product introduced above is the natural indefinite inner product that is invariant under $\text{U}(1, 1)$ transformations. We refer to Ref. \cite{tisseur_2003} for more information about the pseudo-unitary group. The Lie algebra $\mathfrak{u}(1, 1) = \mathrm{span}_\mathbb{R}\{i1_2, iK_x, iK_y, iK_z\}$ is spanned by four pseudo-skew-Hermitian matrices and can be decomposed as $\mathfrak{u}(1, 1) \cong \mathfrak{su}(1, 1) \times \mathfrak{u}(1)$. We use
\begin{align*}
    K_x &= \begin{pmatrix} 1 & 0 \\ 0 & -1 \end{pmatrix}, &
    K_y &= \begin{pmatrix} 0 & 1 \\ -1 & 0 \end{pmatrix}, &
    K_z &= \begin{pmatrix} 0 & i \\ i & 0 \end{pmatrix}, &  
\end{align*}
for $\mathfrak{su}(1, 1)$ and the $2\times 2$ identity matrix $1_2$ for $\mathfrak{u}(1)$. For later use, we compute $\tr(K_j K_k) = 2 g_{jk}$ and $\tr(K_j K_k K_l) = 2i\varepsilon_{jkl}$ where $j, k, l \in \{x, y, z\}$, $g = \text{diag}(+1, -1, -1)$ and $\varepsilon$ is the Levi-Civita symbol. The transformation matrix $T \in \mathrm{U}(1, 1)$ can thus be parameterized by four real numbers using Eq. \eqref{eq:para_unitary_constraint} in the commonly written form:
\begin{equation} \label{eq:transformation_matrix_parameterized}
    T(x, \bm{\vartheta}) = e^{i\vartheta_1}\begin{pmatrix}
        e^{i\vartheta_2} \cosh(x) & e^{i\vartheta_3}\sinh(x)
        \\
        e^{-i\vartheta_3} \sinh(x) & e^{-i\vartheta_2}\cosh(x)
    \end{pmatrix}.
\end{equation}
It follows that bosonic Bogoliubov-Valatin transformations are hyperbolic transformations and that $\mathrm{U}(1, 1)$ is not compact, meaning that entries of $T$ may grow unbounded. In contrast, if the operators in Eq. \eqref{eq:general_hamiltonian} had been fermionic, satisfying $\acomm{\varphi_n}{\varphi_m} = 0$ and $\acomm{\varphi_n}{\varphi_m^\dagger} = \delta_{n m} 1_\mathcal{F}$, we would have $T_{\text{fermion}} \in \text{U}(2)$. The unitary group $\text{U}(2)$ is compact and preserves the Euclidean norm. This distinction between the bosonic and fermionic Bogoliubov-Valatin transformations mirrors the difference between Lorentz and Galilean transformations in special relativity. 

The pseudo-Hermitian dynamical matrix is naturally expressed as a linear combination of the pseudo-Hermitian matrices $1_{2}$ and $\bm{K} = (K_x, K_y, K_z)$ as
\begin{equation} \label{eq:dynamical_matrix}
    D = d_0 1_2 + \bm{d} \cdot \bm{K},
\end{equation}
with $d_0 \in \mathbb{R}$ and the dynamical vector $\bm{d} \in \mathbb{R}^3$ being
\begin{align} \label{eq:d_vec}
    &d_0 = \frac{f-g}{2},  &\bm{d} = \left(\frac{f+g}{2}, \,\, r_1, \,\, r_2\right)^\mathsf{T}. 
\end{align}
The eigenvalues of $D$ are
\begin{equation} \label{eq:dynamical_eigenvalues}
    \lambda_{\pm} = d_0 \pm \norm{\bm{d}}_\text{L},
\end{equation}
where we defined the Lorentzian norm 
\begin{equation} \label{eq:lorentzian_norm}
	\norm{\bm{d}}^2_\text{L} = \bm{d} \cdot_{\scriptscriptstyle \text{L}} \! \bm{d} = d_i g_{ij} d_j = d_x^2 - d_y^2 - d_z^2,
\end{equation}
and repeated indices are summed over. This norm is naturally induced as the Killing form on $\mathfrak{su}(1, 1)$. The excitation energies can now be found through $\sigma_z E = \Lambda = \text{diag}(\lambda_+, \lambda_-)$, which gives $E_\pm = \pm \lambda_\pm$. We note that the sum of excitation energies is proportional to the Lorentzian norm, $\norm{\bm{d}}_\mathrm{L} = (E_+ + E_-) /2$. Therefore, we have $\norm{\bm{d}}_\mathrm{L} > 0$ because the excitation energies were assumed to be positive and real.

\, \\ \subsection{Quantum geometry}
The conventional quantum geometric tensor (QGT) captures the local geometry and topological features of the eigenvectors of the single-particle Hamiltonian as they evolve in parameter space \cite{provost_1980, berry_1984, wilczek_1989}. However, as argued earlier, it is the matrix $D=\sigma_z H$ that governs the dynamics of this system. We therefore expect the relevant quantum geometric tensor for bosonic Bogoliubov Hamiltonians to describe the local geometry and twisting of eigenvectors of $D$, not $H$. To emphasize this distinction from the conventional QGT, we refer to the bosonic Bogoliubov quantum geometric tensor as the symplectic QGT (SQGT) \cite{eckardt_2024}. Note that the non-Hermitian dynamical matrix $D$ has distinct left (L) and right (R) eigenvectors. This allows for the definition of LL, LR, and RR geometric quantities \cite{ma_2022}. Here, we only use the LR quantities as they are the only ones that appear in the transport coefficients to be computed. 

The QGT can be formulated in terms of eigenprojectors of the single-particle Hamiltonian \cite{resta_2011, graf_2021}. We create the SQGT by substituting the projection operator in the conventional QGT with the LR pseudo-orthogonal projection operator $P_{\pm}$ on the eigenspace associated with eigenvalue $\lambda_{\pm}$ of $D$. This yields
\begin{equation} \label{eq:bosonic_qgt}
\begin{split}
    Q_{\mu \nu}^{\pm} &= \tr[\partial_\mu P_{\pm} (1_2 - P_{\pm}) \partial_\nu P_{\pm}].
\end{split}
\end{equation}
The substitution of projection operators is inspired by how Ref. \cite{shindou_2013} first introduced a magnonic Berry curvature relevant for spin wave systems. Reference \cite{eckardt_2024} argues for the relevance of the SQGT, and we will later give more meaning to the SQGT and compare it with the conventional QGT.

The pseudo-orthogonal projection operator $P_\pm$ may be constructed as given in Ref. \cite{shindou_2013},
\begin{equation}\label{eq:bad_projection}
	P_\pm = T \, \Gamma_\pm \, T^\#,
\end{equation}
where $\Gamma_+ = \text{diag}(1, 0)$ and $\Gamma_- = \text{diag}(0, 1)$. The SQGT can subsequently be found by finding $T$ in the form of Eq. \eqref{eq:transformation_matrix_parameterized} and inserting Eq. \eqref{eq:bad_projection} into Eq. \eqref{eq:bosonic_qgt}. However, even for the two-band case, closed-form expressions for $T$ are complicated and may include singularities at certain points in parameter space \cite{owerre_2016}. Additionally, the global phase arbitrariness needs careful consideration for numerical computations involving derivatives. Reference \cite{graf_2021} demonstrates that the conventional QGT can be formulated in an advantageous way for analytical calculations, that is, without constructing eigenvectors. In the following, we adapt the methodology in Ref. \cite{graf_2021} to the SQGT.

We construct $P_\pm$ as the Frobenius covariant \cite{johnson_1991} of the dynamical matrix,
\begin{equation}
    P_{\pm} = \frac{1}{\lambda_{\pm} - \lambda_{\mp}}\left(D - \lambda_{\mp} 1_2\right).
\end{equation}
Inserting $D$ from Eq. \eqref{eq:dynamical_matrix} and $\lambda_{\pm}$ from Eq. \eqref{eq:dynamical_eigenvalues} gives
\begin{equation} \label{eq:dynamical_projection_operator}
    P_{\pm} = \frac{1}{2}(1_2 \pm \hat{\bm{d}} \cdot \bm{K}), 
\end{equation}
where $\hat{\bm{d}} = \bm{d} / \norm{\bm{d}}_\text{L}$ is a Lorentzian-like unit vector. Substituting Eq. \eqref{eq:dynamical_projection_operator} into Eq. \eqref{eq:bosonic_qgt} yields
\begin{equation}
\begin{split}
	Q_{\mu \nu}^\pm &= \frac{1}{8} (\partial_\mu \hat{d}_j) (\partial_\nu \hat{d}_k) \tr \small(K_j K_k \mp \hat{d}_l K_j K_l K_k \small)
	\\
	&= \frac{1}{4} (\partial_\mu \hat{d}_j) (\partial_\nu \hat{d}_k) \small(g_{jk} \mp i \hat{d}_l \varepsilon_{jlk} \small)
	\\
	&= \frac{1}{4} [\partial_\mu \hat{\bm{d}} \cdot_{\scriptscriptstyle \text{L}} \partial_\nu \hat{\bm{d}}  \, \pm \, i \hat{\bm{d}} \cdot (\partial_\mu \hat{\bm{d}} \cross \partial_\nu \hat{\bm{d}} \,)],
\end{split}
\end{equation}
where $j, k, l \in \{x,y,z\}$ and repeated indices are summed over. The SQGT is thus seen to have a simple dependence on the dynamical unit vector and its derivatives in parameter space. The first term is purely real and the second term is purely imaginary. We define the symplectic quantum metric and symplectic Berry curvature in analogy to the conventional quantum metric and Berry curvature \cite{wilczek_1989, resta_2011, eckardt_2024}.

\subsubsection{Quantum metric}
The symplectic quantum metric $G_{\mu \nu}^\pm = \Re Q_{\mu \nu}^\pm$ is
\begin{align} \label{eq:bosonic_quantum_metric_skyrmion}
	G_{\mu \nu}^\pm &= \frac{1}{4} \partial_\mu \hat{\bm{d}} \cdot_{\scriptscriptstyle \text{L}} \! \partial_\nu \hat{\bm{d}}, 
	\\
	&= \frac{1}{4 \norm{\bm{d}}^2_{\text{L}}}\left[\partial_\mu \bm{d} \cdot_{\scriptscriptstyle \text{L}} \partial_\nu \bm{d}  - \frac{(\bm{d} \cdot_{\scriptscriptstyle \text{L}} \partial_\mu \bm{d})(\bm{d} \cdot_{\scriptscriptstyle \text{L}} \partial_\nu \bm{d})}{\norm{\bm{d}}^2_{\text{L}}}\right]. \notag
\end{align}
This expression is similar to the conventional quantum metric in Ref. \cite{graf_2021}, but here it is the dynamical unit vector, and not the Hamiltonian unit vector (Bloch vector), that enters the expression. In addition, the modified scalar product naturally appears. Moreover, we have from Eq. \eqref{eq:d_vec} that $\norm{\bm{d}}_{\text{L}} = (E_+ + E_-)/ 2$. This indicates that the symplectic quantum metric grows where the sum of excitation energies is small, and it does not seem to diverge in regions with crossings or overlaps of the two excitation energies. In contrast, the conventional quantum metric has the difference in excitation energies in its denominator \cite{graf_2021}.

The distance between quantum states may also be formulated in terms of projection operators \cite{bures_1969, resta_2011}. As before, we use the pseudo-orthogonal projection operators that result in
\begin{equation}
	\dd s^2_{\pm}(\bm{k}, \bm{k}^\prime) = 1 - \tr[P_\pm(\bm{k}) P_\pm(\bm{k}^\prime)].
\end{equation}
Now, we let $\bm{k}^\prime = \bm{k} + \dd \bm{k}$ and Taylor expand to second order around $\bm{k}$, which gives
\begin{equation}
	\dd s^2_\pm = \tr[(\partial_\mu P_\pm)(\partial_\nu P_\pm)] \dd k^\mu \dd k^\nu = G_{\mu \nu}^{\pm} \dd k^\mu \dd k^\nu,
\end{equation}
where $P_\pm^2 = P_\pm$ and Eq. \eqref{eq:dynamical_projection_operator} were used. This justifies interpreting $G_{\mu \nu}^\pm$ as a metric.

\subsubsection{Berry curvature}
The symplectic Berry curvature $F_{\mu \nu}^\pm = -2 \Im Q_{\mu \nu}^\pm$ is 
\begin{equation}
\begin{split} \label{eq:pseudo_berry_curvature}
    F_{\mu \nu}^\pm &= \mp \, \frac{1}{2} \, \hat{\bm{d}} \cdot (\partial_\mu \hat{\bm{d}} \cross \partial_\nu \hat{\bm{d}} \, )
    \\
    &= \mp \frac{1}{2\norm{\bm{d}}^3_{\text{L}}} \bm{d} \cdot (\partial_\mu \bm{d} \cross \partial_\nu \bm{d}).
\end{split}
\end{equation}
The magnonic Berry curvature introduced in Ref. \cite{shindou_2013} is identical to our definition of the symplectic Berry curvature as the imaginary part of the SQGT. Therefore, it correctly captures geometric phase evolution in parameter space \cite{shindou_2013, lorenza_2020}. However, in contrast to earlier studies, Eq. \eqref{eq:pseudo_berry_curvature} provides a simple analytical expression for the symplectic Berry curvature in two-band systems. 

Similarly as for the symplectic quantum metric, we find that the symplectic Berry curvature grows where the sum of excitation energies is small. This is in sharp contrast to the conventional Berry curvature that typically diverges in regions with band crossings \cite{graf_2021}. In addition, a nonzero symplectic Berry curvature requires all components of the dynamical vector to be nonzero. Consequently, a nonzero symplectic Berry curvature requires both real and imaginary nonzero off-diagonal terms, along with a nonzero trace, of the single-particle Hamiltonian. Note that the symplectic Berry curvature is expressed with the Euclidean scalar and cross product, whereas the symplectic quantum metric in Eq. \eqref{eq:bosonic_quantum_metric_skyrmion} involved the modified scalar product. This is because Eq. \eqref{eq:pseudo_berry_curvature} corresponds to a signed volume that is independent of the differences in inner product structures. 

\subsubsection{Topological properties} \label{sec:topological_properties}
\begin{figure}[htb]
    \centering
    \includegraphics[width=\linewidth]{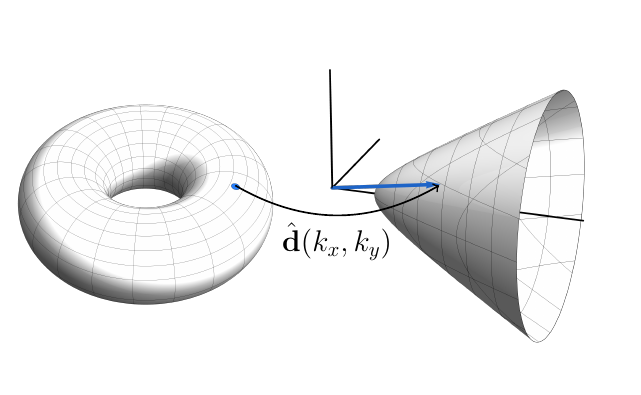}
    \caption{Schematic of the Lorentzian-like dynamical unit vector as a map from parameter space to one of the connected components of a two-sheeted hyperboloid. In this figure, the parameter space is given by $(k_x, k_y)$ that lives on a torus typical for the first magnetic Brillouin zone in two-dimensional periodic systems.}
    \label{fig:torus_hyperboloid}
\end{figure}

We investigate the topological properties using the previously established framework. This can be done by studying the dynamical unit vector $\hat{\bm{d}}$ because the parameter dependence of the projection operators is given through this vector. The Lorentzian norm in Eq. \eqref{eq:lorentzian_norm} indicates that $\hat{\bm{d}}$ lives on a two-sheeted hyperboloid defined by $x^2 - y^2 - z^2 = 1$. However, the positive definite assumption of the single-particle Hamiltonian restricts $x>0$, limiting the space to one of the connected components, which we henceforth refer to as a hyperbolic sheet. In this way we interpret $\hat{\bm{d}}$ as a mapping from the first Brillouin zone to the hyperbolic sheet.
This is illustrated in Fig. \ref{fig:torus_hyperboloid}. 

All such mappings are homotopy equivalent because the hyperbolic sheet is contractible \footnote{We parameterize the hyperbolic sheet by $x = \sqrt{1+r^2}$, $y = r\sin(\phi)$, $z = r\cos(\phi)$ where $r^2 = x^2 + y^2$, $r \geq 0$, $\phi \in [0,2\pi \rangle$. Then, constructing the deformation retract 
\[\bm{H}(r, \phi, t) = \left(\sqrt{1+\left((1 - t)r\right)^2}, (1-t)r\sin(\phi), (1-t)r\cos(\phi)\right), \] that continuously shrinks the hyperbolic sheet to a single point for $t \in [0, 1]$, shows that the hyperbolic sheet is contractible.}. Therefore, the topology associated with the band structure is trivial. This is in agreement with the second no-go theorem in Ref. \cite{lorenza_2020}. Lastly, this result also contrasts with the two-band fermionic Hamiltonian, where the Hamiltonian unit vector (Bloch vector) lives on a two-sphere that is not contractible and allows for nontrivial invariants that classify the topology.

\section{Lieb lattice altermagnet} \label{sec:lieb_lattice_altermagnet}
We now consider a minimal altermagnet model on the Lieb lattice introduced in Refs. \cite{sudbo_2023,maeland_2024}. The altermagnetic properties are here induced by nonmagnetic atoms that make the spins on the two magnetic sublattices related, in the ground state, by a combined spin flip and 90$^\circ$ rotation. Compelling evidence for altermagnetism in the quasi-two-dimensional oxychalcogenides $\mathrm{K V_2 Se_2O}$ \cite{jiang_2024} and $\mathrm{Rb_{1-\delta} V_2 Te_2 O}$ \cite{zhang_2024}, for which our minimal model appears to be a relevant description, has recently been reported using spin-resolved angle-resolved photoemission spectroscopy. Another promising altermagnet candidate on a Lieb lattice is the correlated insulating $d$-wave altermagnet $\mathrm{La_2 O_3 Mn_2 Se_2}$ \cite{AMInsulator_2025}. For this compound, one may expect that spin-splitting of electronic and magnonic bands can be controlled by strain, an effect that has been demonstrated in density functional theory calculations in the altermagnetic semiconductor $\alpha \mathrm{-Mn Te}$ \cite{qauimzadeh_2025}. In particular, for the results on magnon Hall conductivities to be presented in Section \ref{subsec:magnon_hall_conductivities}, $\mathrm{La_2 O_3 Mn_2 Se_2}$  is a promising platform, since the conductivities could be measured under varying strains. 

On top of the spin Hamiltonian in Ref. \cite{sudbo_2023}, we add a Dzyaloshinskii-Moriya interaction (DMI). In the present case, this is necessary to induce a nonzero Berry curvature and hence nonzero Hall conductivities. Physically, DMI can be induced in such systems by growing the layer on top of a heavy metal or sandwiching it between two different layers \cite{brataas_2018}. A DMI can be interpreted as a gauge field that induces complex hopping terms with phase factors \cite{hotta_2019}, which is the underlying reason for the emergence of a nonzero imaginary term in the single-particle Hamiltonian in this model.
\begin{figure}[htb]
    \centering
    \includegraphics[width=\linewidth]{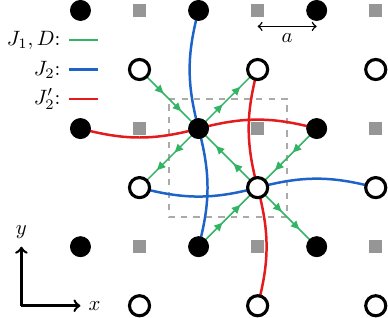}
    \caption{A section of a two-dimensional Lieb lattice. The magnetic sublattices $L_1$ and $L_2$ are denoted by filled and hollow circles, respectively. nonmagnetic atoms are shown as squares, the lattice spacing is $a$, and the chosen unit cell is marked by the dashed square. The couplings involving at least one atom in the unit cell depicted are shown by the colored lines.}
    \label{fig:couplings}
\end{figure}

The two-dimensional Lieb lattice is illustrated in Fig. \ref{fig:couplings}. We divide the magnetic atoms into two magnetic sublattices, $L_1$ and $L_2$. In the ground state, spins on $L_1$ $(L_2)$ point in the positive (negative) $z$-direction. The nonmagnetic atoms are responsible for real-space direction-dependent second-nearest neighbor couplings \cite{sudbo_2023}. It is the difference between the two second-nearest neighbor couplings, which we name the altermagnetic parameter, that governs the strength of the altermagnetic splitting. The nonmagnetic atoms also allow for nearest neighbor intrinsic DMI due to the absence of an inversion center between magnetic atoms on different magnetic sublattices.

\, \\ The spin Hamiltonian used to describe the low-energy spin wave excitations is
\begin{equation}
\begin{split}
    \mathcal{H} = &\sum_{i, j} [J_{ij}  \bm{S}_i \cdot \bm{S}_j + D_{ij} \hat{z} \cdot (\bm{S}_i \cross \bm{S}_j)] 
    \\
    - &\sum_i [K (S^z_i)^2 + B  S_i^z ],
\end{split}
\end{equation}
where the summation over $i$ and $j$ runs over all magnetic atoms, $\bm{S} = (S^x, S^y, S^z)$ are spin operators, $K$ accounts for easy-axis anisotropy and $B$ is the strength of an external magnetic field. We consider exchange interactions up to and including second-nearest neighbors. The nearest neighbor Heisenberg exchange interaction $J_1 > 0$ is responsible for antiferromagnetic order. The second-nearest neighbor coupling $J_2 < 0$ is a direct coupling, while $J^\prime_2 < 0$ is mediated by a nonmagnetic atom, which is illustrated in Fig. \ref{fig:couplings}. Following Moriya's rules \cite{moriya_1960}, we deduce that the nearest neighbor DMI must be out of plane. We choose its sign as determined by the direction of the arrows in Fig. \ref{fig:couplings}: $D_{ij} = +D$ if $i \to j$ follows the arrows and $D_{ij} = -D$ if it goes against them. When $J^\prime_2 = J_2$, the model reduces to the $J_1-J_2$ Heisenberg model on the square lattice with easy-axis anisotropy, magnetic field, and nearest neighbor DMI. The relative strength of the different parameters is chosen to reflect a ground state with Nèel order as in Ref. \cite{sudbo_2023}.

We apply the Holstein-Primakoff transformation \cite{primakoff_1940} to describe spin wave excitations, which to leading order in bosonic operators are
\begin{subequations}
\begin{align}
    &\qq{if} i \in L_1: \left\{
    \begin{array}{ll}
        S^+_i &\approx \sqrt{2S} \, b_{1, i},
        \\
        S^-_i &\approx \sqrt{2S} \, b_{1,i}^\dagger, \\
        S^z_i &= S - b_{1,i}^\dagger b_{1,i},
    \end{array}
    \right.
    \\
    &\qq{if} i \in L_2: \left\{
    \begin{array}{ll}
        S^+_i &\approx \sqrt{2S} \, b_{2,i}^\dagger, \\
        S^-_i &\approx \sqrt{2S} \, b_{2,i}, \\
        S^z_i &= b_{2,i}^\dagger b_{2,i} - S,
    \end{array}
    \right.
\end{align}
\end{subequations}
where $S^\pm = S^x \pm i S^y$, $b_1$ and $b_2$ are bosonic operators referring to magnetic sublattices $L_1$ and $L_2$, and $S$ is the magnitude of the localized spins. Applying the Fourier transforms
\begin{align*}
    &b_{1,i} = \frac{1}{\sqrt{V}}\sum_{\bm{k}} e^{-i \bm{k} \cdot \bm{r}_i} \, b_{1, \bm{k}}, &b_{2,i} = \frac{1}{\sqrt{V}}\sum_{\bm{k}} e^{-i \bm{k} \cdot \bm{r}_i} \, b_{2, \bm{k}},
\end{align*}
where $V$ is the number of unit cells, leads to
\begin{equation} \label{eq:two_band_hamiltonian_k}
    \mathcal{H} = \frac{1}{V} \sum_{\bm{k}} 
    \begin{pmatrix} 
    b^\dagger_{1, \bm{k}} & b_{2, -\bm{k}}
    \end{pmatrix}
    \begin{pmatrix}
        f_{\bm{k}} & r_{\bm{k}}
        \\
        r^*_{\bm{k}} & g_{\bm{k}}
    \end{pmatrix}
    \begin{pmatrix} b_{1, \bm{k}} \\ b^\dagger_{2, -\bm{k}}\end{pmatrix}.
\end{equation}
This takes the form of the two-band bosonic Bogoliubov Hamiltonian in Eq. \eqref{eq:general_hamiltonian}, where $r_{\bm{k}} = r_{1,\bm{k}} + ir_{2,\bm{k}}$ and
\begin{subequations}
\begin{align}
    f_{\bm{k}} &= 2 S J^\prime_2 \cos(2k_x a) + 2S J_2 \cos(2k_y a)  
    \\
    &- 2S(J_2 + J^\prime_2) + 4SJ_1 + 2SK + B, \notag
    \\[0.2cm]
    g_{\bm{k}} &= 2 S J^\prime_2 \cos(2k_y a) + 2SJ_2 \cos(2k_x a)
    \\
    &- 2S(J_2 + J^\prime_2) + 4SJ_1 + 2SK - B, \notag
    \\[0.2cm]
    r_{1,\bm{k}} &= 2 S J_1 [\cos(k_x a - k_y a) + \cos(k_x a + k_y a)],
    \\[0.2cm]
    r_{2, \bm{k}} &= 2S D [\cos(k_x a - k_y a) - \cos(k_x a + k_y a)].
\end{align}
\end{subequations}

\subsection{Magnon bands and Berry curvature}
\begin{figure}[htb]
    \centering
    \includegraphics[width=\linewidth]{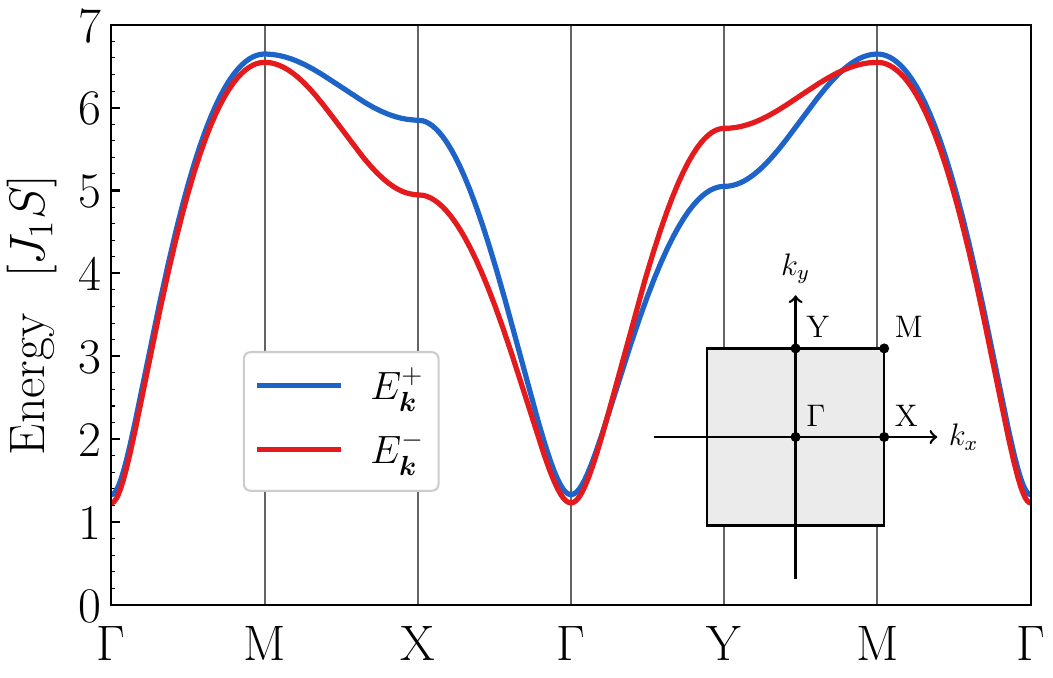}
    \caption{Magnon excitation energies for the Lieb lattice altermagnet with parameter values given in table \ref{table:parameters}. The first magnetic Brillouin zone is represented by the filled square where the high symmetry points in momentum space are $\Gamma = (0, 0)$, X $= (\pi / 2a, 0)$, M $ = (\pi / 2a, \pi / 2a)$, Y $=(0, \pi / 2a)$.}
    \label{fig:magnon_bands}
\end{figure}

The excitation energies are found as outlined in Eqs. \eqref{eq:dynamical_eigenvalues} and \eqref{eq:lorentzian_norm}. We define the sum of excitation energies $\omega = (E_+ + E_-) / 2$ and the difference $\delta = (E_+ - E_-)/2$, where
\begin{subequations} \label{eq:omega_delta}
\begin{align} 
	\frac{\omega_{\bm{k}}^2}{S^2} &= \{(J^\prime_2 + J_2) [\cos(2k_x a) + \cos(2k_y a) - 2]  + 4J_1 + 2 K\}^2 \notag
	\\
	& - 16 J_{1}^2 \cos^2(k_x a) \cos^2(k_y a) -4 D^2 \sin^{2}(k_x a) \sin^{2}(k_y a), \notag
	\\[0.2cm]
	\frac{\delta_{\bm{k}}}{S} &= (J^\prime_2 - J_2)[\cos(2k_x a) - \cos(2k_y a)] + \frac{B}{S}. \label{eq:delta} 
\end{align}
\end{subequations}
The difference between excitation energies renders the altermagnetic character, which is a $d$-wave $(d_{x^2 - y^2})$ symmetry in momentum space for $B = 0$. It is the difference between the two next-nearest neighbors couplings $|J^\prime_2 - J_2| \neq 0$ named the altermagnetic parameter that governs the strength of the altermagnetic splitting. Recently, it has been shown that strain fields can tune such a parameter \cite{qauimzadeh_2025}. For later use, we notice that both $\omega$ and $\delta|_{B = 0}$ are even under $\bm{k} \to -\bm{k}$, whereas $\omega$ is even and $\delta|_{B=0}$ is odd under a 90$^\circ$ rotation in momentum space. The excitation energies are shown in Fig. \ref{fig:magnon_bands} with parameters given in table \ref{table:parameters}. 

\begin{figure}[htb]
    \centering
    \includegraphics[width=\linewidth]{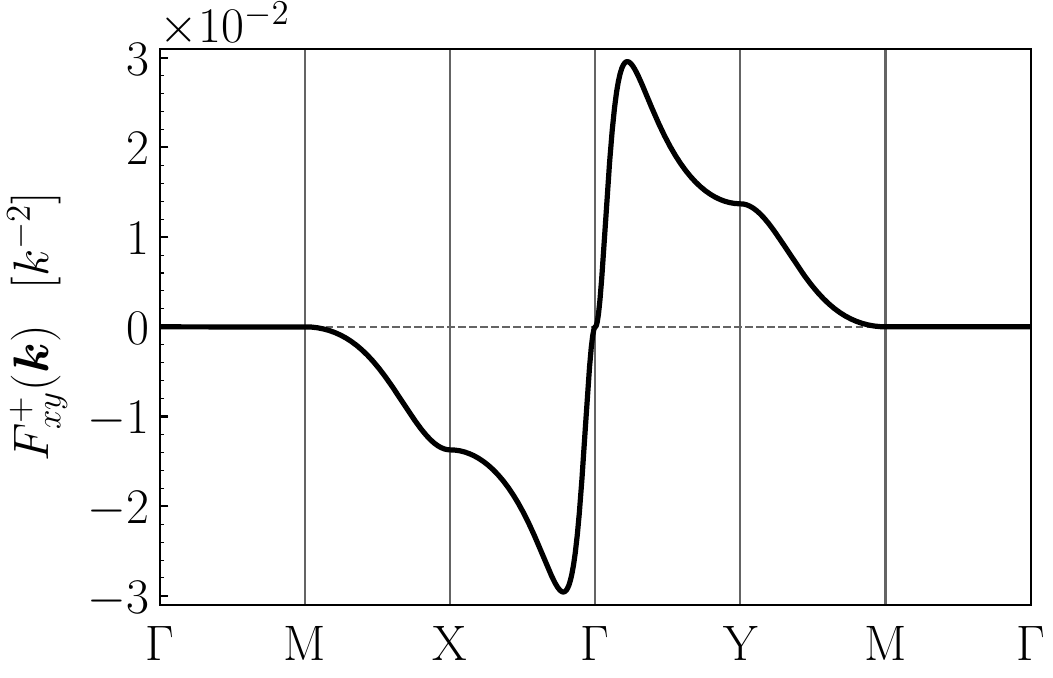}
    \caption{Berry curvature for the Lieb lattice altermagnet. The parameter values are given in table \ref{table:parameters} and the high symmetry points in momentum space are given in Fig. \ref{fig:magnon_bands}.}
    \label{fig:berry}
\end{figure}

The Berry curvature is computed using Eq. \eqref{eq:pseudo_berry_curvature} with the result
\begin{align}  \label{eq:bosonic_berry_curvature_end}
	&F^\pm_{xy}(\bm{k}) = \pm D J_1 \omega_{\bm{k}}^{-3} S^3 a^2 \notag
	\\
	\times &\{ \left[8J_1 - 4(J_2 +J^\prime_2) + 4K\right] \left[\cos(2k_x a) - \cos(2k_y a) \right] \notag
	\\
	&\hspace{0.3cm}-\left(J_2 + J^\prime_2\right) \left[\cos(4 k_x a) - \cos(4 k_y a) \right] \}.
\end{align}
This quantity is shown in  Fig. \ref{fig:berry} along high-symmetry directions in the first magnetic Brillouin zone. It is seen that $F_{xy}^{\pm}(\bm{k})$ is even under $\bm{k} \to -\bm{k}$ and odd under a 90$^\circ$ rotation in momentum space. Therefore, the momentum space integral over the first magnetic Brillouin zone, proportional to the first Chern number \cite{shindou_2013}, is zero. The topologically trivial band structure agrees with the analysis performed in Sec. \ref{sec:topological_properties}. In addition, the analytic expression shows that the Berry curvature vanishes when $D=0$ as expected. Moreover, it is independent of the magnetic field and only the sum, and not the difference, of the second-nearest neighbor couplings appears in the expression. The Berry curvature is therefore insensitive to the strength of the altermagnetic splitting.

\begin{table}[htb] 
    \centering
    \caption{Spin Hamiltonian parameters used in this study, unless otherwise specified.}
    \begin{tabular}{c|c|c|c|c|c|}
         \text{Parameter} & $J_2 / J_1$ & $J^\prime_2 / J_1$ & $K / J_1$ & $D / J_1$ & $B / J_1 S$\\
         \hline 
         \text{Value} & -0.4 & -0.2 & 0.1 & 0.1 & 0.05
    \end{tabular} \label{table:parameters}
\end{table}

\subsection{Magnon Hall conductivities}
\label{subsec:magnon_hall_conductivities}
The magnon thermal Hall conductivity $\kappa$ and the magnon spin Nernst conductivity $\alpha$ describe respectively a transverse heat and spin current due to an applied longitudinal temperature gradient. The conductivities were in Refs.  \cite{murakami_2014, kovalev_2016} both derived within quantum linear response theory for Hamiltonians satisfying particle-hole symmetry \cite{kondo_2020}, which Eq. \eqref{eq:two_band_hamiltonian_k} does not. Up to an additive constant, Eq. \eqref{eq:two_band_hamiltonian_k} can be rewritten in a particle-hole symmetric form by doubling the number of operators in the Nambu array \cite{colpa_1978}. Then, Ref. \cite{pires_2023} shows that employing the particle-hole symmetry results in
\begin{subequations} \label{eq:hall_conductivities}
\begin{align}
    \alpha_{xy} &= -\frac{k_\text{B}}{\hbar V} \sum_{\bm{k}}\{c_1[n_{\text{B}}(E^+_{\bm{k}})] - c_1[n_{\text{B}}(E^-_{-\bm{k}})] \}F^+_{xy}(\bm{k}), 
    \\
    \kappa_{xy} &= -\frac{k_\text{B}^2 T}{\hbar V} \sum_{\bm{k}}\{c_2[n_{\text{B}}(E^+_{\bm{k}})] + c_2[n_{\text{B}}(E^-_{-\bm{k}})]\} F^+_{xy}(\bm{k}), 
\end{align}
\end{subequations}
where $k_\mathrm{B}$ is the Boltzmann constant, $\hbar$ is Planck's reduced constant, $n_B(x) = [\exp(\beta x) - 1]^{-1}$ is the Bose-Einstein distribution, $\beta = 1 / (k_\mathrm{B} T)$ is the inverse temperature, and
\begin{subequations}
\begin{align}
    c_1(x) &= (1 + x)\ln(1 + x) - x\ln(x),
    \\
    c_2(x) &= (1 + x)\left[\ln\left(\frac{1 + x}{2}\right)\right]^2 - [\ln(x)]^2 - 2\text{Li}_2(-x),
\end{align}
\end{subequations}
with the polylogarithm function being $\text{Li}_2(x) = \sum_{k = 1}^{\infty} x^{k} / k^{2}$. The sign difference between the terms involving the distribution functions in Eq. \eqref{eq:hall_conductivities} arises because of the differences between spin and heat currents \cite{pires_2023}, where the two magnon bands carry opposite $z$-components of spin. A heat current changes sign under time reversal, while a spin current and temperature gradient do not. Therefore, $\kappa \neq 0$ requires broken time reversal symmetry, but $\alpha \neq 0$ does not. The ground state has an effective time reversal symmetry given by a combined spin flip and 90$^\circ$ rotation, such that in this model it is the presence of a nonzero magnetic field that allows $\kappa \neq 0$. 

\begin{figure}[htb]
    \centering
    \includegraphics[width=\linewidth]{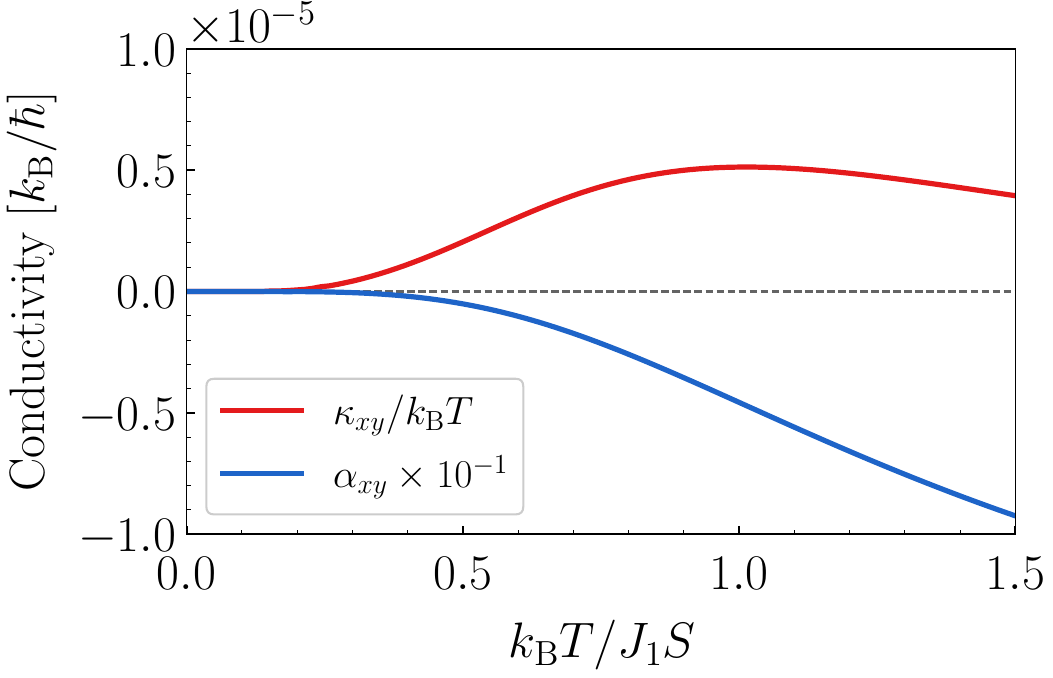}
    \caption{Magnon Hall conductivities as a function of temperature in the Lieb lattice altermagnet. The parameter values used are given in table \ref{table:parameters}.}
    \label{fig:conductivity_t}
\end{figure}

We compute the integrand in Eq. \eqref{eq:hall_conductivities} symbolically using Eq. \eqref{eq:omega_delta} for the excitation energies and Eq. \eqref{eq:bosonic_berry_curvature_end} for the Berry curvature. An adaptive numerical integration method is used to compute the summation over reciprocal lattice vectors in the thermodynamic limit. The magnon Hall conductivities are shown in Fig. \ref{fig:conductivity_t} as a function of temperature.

\subsubsection{High and low temperature limit}
\begin{figure}[htb]
    \centering
    \includegraphics[width=\linewidth]{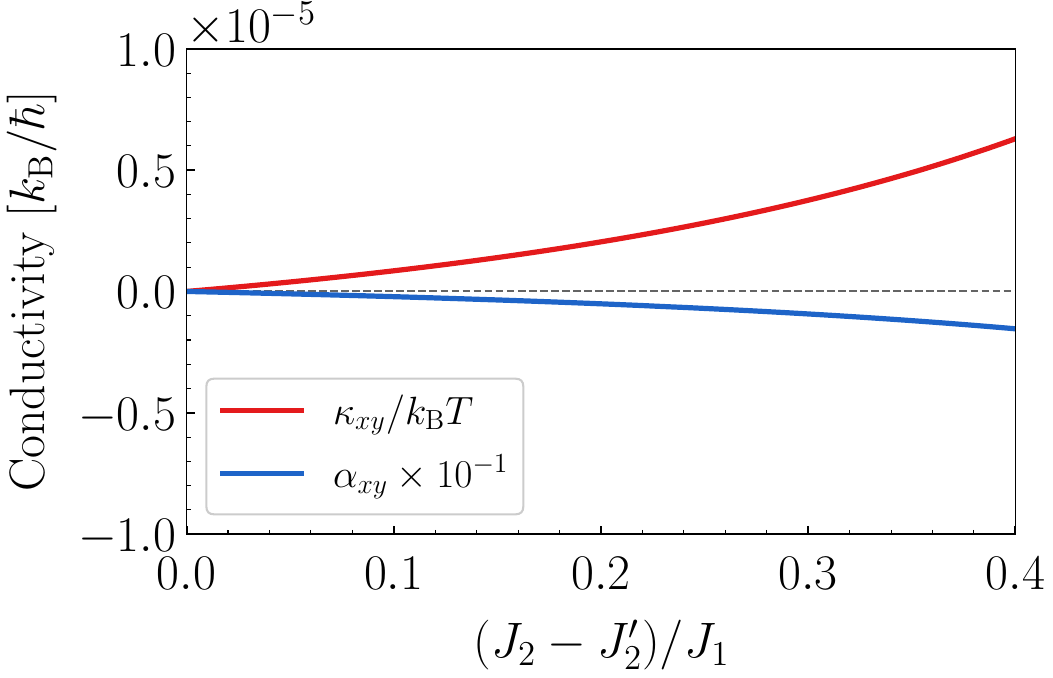}
    \caption{Magnon Hall conductivities as a function of the altermagnetic strength $(J_2-J_2^{\prime})/J_1$ with $k_\mathrm{B} T / J_1 S = 0.5$ and the other parameter values as given in table \ref{table:parameters}. A standard square lattice antiferromagnet is recovered when $J^\prime_2=J_2$.}
    \label{fig:conductivity_am}
\end{figure}

We next \textit{}analyze the high and low temperature limits of the conductivities. These can be understood from the asymptotic expansions of $c_1$ and $c_2$:
\begin{subequations}
\begin{align}
	c_1(x)  &\underset{x \ll 1}{\approx} -x\ln(x),& c_1(x)  &\underset{x \gg 1}{\approx}  \ln(x),
	\\
	c_2(x)  &\underset{x \ll 1}{\approx} x\ln^2(x),& c_2(x)  &\underset{x \gg 1}{\approx} \pi^{2} / 3 - 1/x + 1 / 2x^2.
\end{align}
\end{subequations}
We use $n_\mathrm{B}(x) \approx \exp(-\beta x)$ at low temperatures, which together with the symmetries in momentum space of the relevant quantities, gives
\begin{subequations}
\begin{align}
    &\alpha_{xy}|_{B = 0} \underset{\mathrm{\beta \gg 1}}{\approx} -\frac{2 \beta k_\mathrm{B}}{\hbar V} \sum_{\bm{k}} F^+_{xy} \, e^{-\beta \omega}
    \\
    & \times [\delta |_{B = 0} \cosh(\beta \delta |_{B = 0})  - \omega \sinh(\beta \delta |_{B = 0})], \notag
    \\[0.2cm]
    &\kappa_{xy} \underset{\mathrm{\beta \gg 1}}{\approx} -\frac{2 \beta k_\mathrm{B}}{\hbar V} \sum_{\bm{k}}  F^+_{xy} \, e^{-\beta \omega}
    \\
    &\times \Big\{2 \delta|_{B=0} \cosh(\beta \delta|_{B=0}) \left[B \cosh(\beta B) - \omega \sinh(\beta B) \right] \notag
    \\
    &\hspace{0.5cm}+ \sinh(\beta \delta|_{B=0}) \big[(\omega^2 + \delta|_{B=0}^{2} + B^2) \sinh(\beta B) \notag
    \\
    &\hspace{3cm}- 2 \sinh(\beta \delta|_{B=0}) \omega B \cosh(\beta B)\big] \Big\}.\notag
\end{align}
\end{subequations}
The conductivities are exponentially suppressed at low temperatures because the sum of excitation energies is much larger than the difference, $\omega / \delta \gg 1$. We also explicitly see that $\kappa_{xy} \neq 0$ requires $B \neq 0$ at low temperatures, while $\alpha_{xy} \neq 0$ does not. At high temperatures, we use $n_\mathrm{B}(x) \approx 1 / (\beta x)$, which gives
\begin{subequations}
\begin{align}
	\alpha_{xy}|_{B = 0} &\underset{\mathrm{\beta \ll 1}}{\approx} -\frac{k_\mathrm{B}}{\hbar V} \sum_{\bm{k}} \ln(\frac{\omega - \delta|_{B=0}}{\omega + \delta|_{B=0}}) F_{xy}^{+},
	\\
	\kappa_{xy} &\underset{\mathrm{\beta \ll 1}}{\approx} -\frac{2B \beta^{2} k_\mathrm{B}}{\hbar V} \sum_{\bm{k}} \delta|_{B=0} \, F_{xy}^{+}.
\end{align}
\end{subequations}
Therefore, as the temperature increases, $\alpha_{xy}$ approaches a constant and $\kappa_{xy}$ approaches zero, which agrees with the trend in Fig. \ref{fig:conductivity_t}. Both conductivities vanish in the low and high temperature limits when $\delta|_{B=0} = 0$, which corresponds to the absence of altermagnetic splitting. It is interesting to investigate if this also holds for moderate temperatures on the scale of the excitation energies. Figure \ref{fig:conductivity_am} shows the conductivities as a function of the difference between the two next-nearest neighbor couplings that control the strength of the altermagnetic splitting for the moderate temperature $k_\mathrm{B} T / J_1 S = 0.5$.

\subsubsection{Zero altermagnetic splitting limit}
Figure \ref{fig:conductivity_am} shows that the conductivities vanish together with the strength of the altermagnetic splitting. We now demonstrate this analytically for all temperatures. The difference in excitation energies, denoted by $\delta$ in Eq. \eqref{eq:delta}, is tunably small in the strength of the altermagnetic splitting and magnetic field. Therefore, we expand the conductivities in Eq. \eqref{eq:hall_conductivities} around $\omega$, which to second order in $\delta$ gives
\begin{subequations}\label{eq:final_expr}
\begin{align} 
    \alpha_{xy} &\underset{\delta \ll 1}{\approx} (J^\prime_2 - J_2)  C^\alpha_{xy} \, 2S k_\text{B}/ \hbar,
    \\[0.2cm]
    \kappa_{xy} &\underset{\delta \ll 1}{\approx} B(J^\prime_2 - J_2) C^\kappa_{xy} \, 2S k_\text{B}^2 T  / \hbar,
\end{align}
\end{subequations}
with
\begin{subequations}
\begin{align}
    C^\alpha_{xy} &= \frac{1}{V}\sum_{\bm{k}} [\cos(2k_x a) - \cos(2k_y a)] p^\prime_1(\omega_{\bm{k}}) F^+_{xy}(\bm{k}), \notag
    \\
    C^\kappa_{xy} &= \frac{1}{V}\sum_{\bm{k}} [\cos(2k_x a) - \cos(2k_y a)] p^{\prime \prime}_2(\omega_{\bm{k}}) F^+_{xy}(\bm{k}). \notag
\end{align}
\end{subequations}
Here, $p_1(x) = c_1[n_\mathrm{B}(x)]$, $p_2(x) = c_2[n_\mathrm{B}(x)]$, and the prime symbol denotes a derivative as $p^\prime_1 = \partial_x\{c_1[n_\mathrm{B}(x)]\}$. Equation \eqref{eq:final_expr} demonstrates that nonzero magnon thermal Hall and spin Nernst conductivities require a finite altermagnetic splitting. This result is consistent with what is known for the square lattice antiferromagnet \cite{nagaosa_2010}. Reference \cite{mook_2025} also calculated magnon Hall conductivities in a model of an altermagnet. However, the altermagnet model of Ref. \cite{mook_2025} differs in a fundamental way from the model used in the present paper, since Ref. \cite{mook_2025} only considers DMI vectors confined to the $xy$-plane for spin ordering along the $z$-axis. This yields $\alpha_{xy} = \kappa_{xy} = 0$ for any values of parameters \cite{mook_2025}. 

\section{Summary} \label{summary}
We have presented a framework for computing the relevant quantum geometric tensor in systems described by two-band bosonic Bogoliubov Hamiltonians. By identifying bosonic Bogoliubov-Valatin transformations as elements of the pseudo-unitary group, we showed that the relevant quantum geometric tensor had a simple dependence on the coefficients in the Hamiltonian. This provided analytical insight into the transport properties. Applying this formalism to a minimal model of an altermagnet, we demonstrated that the magnon thermal Hall and spin Nernst conductivities are directly sensitive to the altermagnetic parameter. The altermagnetic parameter and hence non-relativistic band splitting can be tuned externally by strain fields. We have suggested candidate materials where this could be the case, in particular the altermagnetic insulator $\mathrm{La_2O_3Mn_2Se_2}$. This may offer a potential route for the experimental identification of altermagnetic properties in magnetic insulators based on quantum transport probes. 

\begin{acknowledgments}
E.S. thanks Olav H. Haaland, Bj{\o}rnar G. Hem, and Olav F. Sylju{\aa}sen for useful discussions. We acknowledge support from the Norwegian Research Council through Grant No. 262633, “Center of Excellence on Quantum Spintronics”, and Grant No. 323766, as well as COST Action CA21144  ``Superconducting Nanodevices and Quantum Materials for Coherent Manipulation".
\end{acknowledgments}
\bibliography{references}
\end{document}